\documentclass[aps,onecolumn,floatfix,showpacs]{revtex4}
\usepackage{graphicx}
\usepackage{array}
\usepackage{amsmath}
\usepackage{amssymb}
\usepackage{bm}

\newcommand{\Pf}{\mathrm{Pf}}
\newcommand{\sign}{\mathrm{sign}}
\newcommand{\be}{\mathbf{e}}
\newcommand{\balpha}{\boldsymbol{\alpha}}
\newcommand{\bx}{\mathbf{x}}
\newcommand{\bu}{\mathbf{u}}
\newcommand{\bv}{\mathbf{v}}
\newcommand{\bw}{\mathbf{w}}

\begin{document}

\title{Efficient numerical computation of the Pfaffian for dense and banded
skew-symmetric matrices}

\author{M. Wimmer}
\affiliation{Instituut-Lorentz, Universiteit Leiden, P.O. Box 9506, 2300 RA Leiden, The Netherlands}

\begin{abstract}
Computing the Pfaffian of a skew-symmetric matrix is a problem that arises
in various fields of physics. Both computing the Pfaffian and
a related problem, computing the canonical form of a skew-symmetric matrix
under unitary congruence, can be solved easily once the skew-symmetric
matrix has been reduced to skew-symmetric tridiagonal form. We develop
efficient numerical methods for computing this tridiagonal form based on
Gauss transformations, using a skew-symmetric, blocked form of the
Parlett-Reid algorithm, or based on unitary transformations, using block
Householder transformations and Givens rotations, that are applicable
to dense and banded matrices, respectively. We also give a complete
and fully optimized implementation of these algorithms in Fortran, and
also provide Python, Matlab and Mathematica implementations for convenience.
Finally, we apply these methods to compute the topological
charge of a class D nanowire, and show numerically the equivalence of
definitions based on the Hamiltonian and the scattering matrix.
\end{abstract}


\maketitle

\section{Introduction}
\label{intro}

\subsection{Pfaffians and reduction to tridiagonal form}

A real or complex matrix $A$ is called \emph{skew-symmetric} (or
\emph{anti-symmetric}), if $A=-A^T$, where $^T$ denotes the
transpose. The determinant $\det(A)$ of such a skew-symmetric matrix
is the square of a polynomial in the matrix entries, the
\emph{Pfaffian} $\Pf(A)$:
\begin{equation}
\det(A)=\Pf(A)^2\,.
\end{equation}
In other words, the Pfaffian of a skew-symmetric matrix is a unique choice
for the sign of the root of the determinant:

\begin{equation}
\Pf(A)=\pm \sqrt{\det(A)}
\end{equation}
Pfaffians arise in various fields of physics, such as in the
definition of topological charges \cite{Kitaev2001,Fu2007a,Fulga2011},
electronic structure quantum Monte Carlo \cite{Bajdich2009}, the
two-dimensional Ising spin glass \cite{Thomas2009}, or in the
definition of a trial wave function for the $\nu=5/2$ fractional
quantum Hall state \cite{Moore1991}.  It also arises naturally from
Gaussian Grassmann integration, and as such finds applications for
example in quantum chaos \cite{Haake2004} or lattice quantum field
theory \cite{Montvay1996}.

The Pfaffian for a $2n\times 2n$ skew-symmetric matrix is defined as
\begin{equation}\label{pfaff_def}
\mathrm{Pf}(A)=\frac{1}{2^{n} n!} \sum_{\sigma \in S_{2n}} \mathrm{sgn}(\sigma)
\prod_{i}^{n} a_{\sigma(2i-1),\sigma(2i)}\,
\end{equation}
where $S_{2n}$ is the group of permutations of sets with $2n$ elements. The
Pfaffian of an odd-dimensional matrix is defined to be zero, as
in this case also $\det(A)=0$ ($\det(A)=\det(A^T)=\det(-A)=(-1)^{2n-1} \det(A)$).
While Eq.\ \eqref{pfaff_def} can be used to compute the Pfaffian directly
for small matrices, its computational cost $\mathcal{O}(n!)$ is
prohibitively expensive for larger matrices.

Analogous to the numeric computation of the determinant, a promising strategy
is thus to find a transformation of the original matrix into a form
that allows an easier evaluation of the Pfaffian. Particularly useful
in this context is the recursive definition of the Pfaffian,
\begin{equation}\label{pfaff_recursive}
\Pf(A)=\sum_{i=2}^{2n} (-1)^i a_{1i} \Pf(A_{1i})\,,
\end{equation}
where $A_{1i}$ is the matrix A without the rows and columns 1 and $i$. (Note
that the Pfaffian of a $0\times 0$ matrix is defined as 1). Further,
for an arbitrary $2n\times 2n$ real or complex matrix $B$,
\begin{equation}\label{pfaff_transform}
\Pf(B A B^T)=\det(B) \Pf(A)\,.
\end{equation}

From the recursive definition of the Pfaffian \eqref{pfaff_recursive} it
is obvious that the Pfaffian of a $2n\times 2n$ skew-symmetric tridiagonal
matrix
\begin{equation}\label{td_form1}
T=\begin{pmatrix}0&a_{1}\\
-a_{1}&0&b_{1}\\
&-b_{1}&0&a_{2}\\
&&-a_2&\ddots&\ddots\\
&&&\ddots&0&b_{n-1}\\
&&&&-b_{n-1}&0&a_{n}\\
&&&&&-a_{n}&0
\end{pmatrix}
\end{equation}
is given as
\begin{equation}
\Pf(T)=\prod_{i=1}^{n} a_i\,.
\end{equation}
Furthermore, a closer inspection of Eq.\ \eqref{pfaff_recursive}
shows that also a matrix that has only a partial tridiagonal form
with $t_{ij}=t_{ji}=0$ only for odd $i$ and $j>i+1$
(i.e. a matrix that would be tridiagonal, if every \emph{even} row and column
would be removed),
\begin{equation}\label{partialT}
\tilde{T}=\begin{pmatrix}0&a_{1}\\
-a_{1}&0&t_{23}&t_{24}&t_{25}&\dots\\
&-t_{23}&0&a_{2}\\
&-t_{24}&-a_2&\ddots&\ddots\\
&-t_{25}&&\ddots&0&t_{2n-2,2n-1}&t_{2n-2,2n}\\
&\vdots&&&-t_{2n-2,2n-1}&0&a_{n}\\
&&&&-t_{2n-2,2n}&-a_{n}&0
\end{pmatrix}
\end{equation}
allows for an easy evaluation of the Pfaffian, as $\Pf(\tilde{T})=\Pf(T)$.
Our goal is therefore to find for a skew-symmetric matrix $A$
a suitable transformation $B$ such that
\begin{equation}\label{pfaff_BTB}
A=B T B^T
\end{equation}
with $T$ tridiagonal or tridiagonal in every odd row and
column.

It has been known for a while that the Pfaffian of a skew-symmetric $n
\times n$ matrix $A$ can be computed in $\mathcal{O}(n^3)$ time, using
a skew-symmetric form of Gaussian elimination (adding multiples of
rows and columns in a symmetric fashion)
\cite{Galbiati1994,Rote2001,Krauth2006,Bajdich2009}.  Such an
skew-symmetric Gaussian elimination computes a factorization of the
matrix in the form \eqref{pfaff_BTB} with $B=P L$ where $P$ is a
permutation matrix and $L$ a unit lower triangular matrix. For
brevity, we will refer to this type of decomposition as $LTL^T$
decomposition. Gaussian elimination requires pivoting for numerical
stability, hence the need for the permutation $P$. Below, we will
formulate this approach in a way that allows for an efficient computer
implementation.

Another Gaussian based elimination technique is the $LDL^T$
decomposition where $A$ is reduced to $D$, a matrix with only
skew-symmetric $2\times 2$-blocks on the diagonal
\cite{Bunch1982,Higham2002}. This approach has also been suggested for
computing the Pfaffian recently \cite{Thomas2009,Rubow2011}.  We will
not persue this approach here, but show that the $LTL^T$ decomposition
allows for computing the Pfaffian in the same number of operations and
can be formulated more easily to use level-3 matrix operations.

As an alternative to the Gaussian elimination based techniques, we also
develop algorithms using unitary (orthogonal in the real case)
transformations that are also known to allow for a stable numerical
computation in $\mathcal{O}(n^3)$ for dense matrices. This approach
doe not require pivoting for numerical stability and can more easily
make use of the bandedness of a matrix. We will describe how to
compute a unitary matrix $Q$ such in order to tridiagonalize (either
fully or partially) $A$,
\begin{equation}
A=Q T Q^T\,,
\end{equation}
or equivalently $T=Q^\dagger A Q^*$, where $^\dagger$ denotes the
Hermitian conjugate and $^*$ complex conjugation. Note that such a
\emph{unitary congruence transformation} is for the complex case quite
different from the usual unitary similarity transformations usually
encounters, which are of the form $A=Q T Q^\dagger$.  In the real
case, the transformation reduces to the usual orthogonal similarity
transformation.

\subsection{Tridiagonalization and the canonical form of skew-symmetric matrices}
\label{Youla}

Apart from computing allowing for an efficient computation of the
Pfaffian, the tridiagonal form of a skew-symmetric matrix under
unitary congruence is also relevant for computing the canonical form
of this matrix.

A skew-symmetric matrix has a particularly simple canonical
form under a unitary congruence transformation. For every
skew-symmetric matrix $A$ there exists
a unitary matrix $U$ such that \cite{Hua1944,Stander1960}
\begin{equation}\label{youla}
A=U \Xi U^T, \quad\text{where $\Xi=\Sigma_1 \oplus \Sigma_2 \oplus
  \dots \oplus \Sigma_k \oplus 0 \oplus \dots \oplus 0$}
\end{equation}
where $\mathrm{rank}(A)=2k$, $\oplus$ denotes the direct sum, and
\begin{equation}
\Sigma_j=\begin{pmatrix}
0&\sigma_j\\
-\sigma_j&0
\end{pmatrix}\,,\qquad \sigma_j>0.
\end{equation}
This canonical form has been used in the physics context for example
to prove the Kramer's degeneracy of transmission eigenvalues
\cite{Bardarson2008} and the degeneracy of Andreev reflection eigenvalues
\cite{Beri2009}.

The problem of computing the canonical form of an even-dimensional
skew-symmetric tridiagonal matrix has been discussed in
\cite{Ward1978a,Golub1996,Hastings2010}, the reduction of
the problem with on odd-dimensional matrix to the even-dimensional
case in \cite{Ward1978a}. For a $2n\times 2n$ skew-symmetric
tridiagonal matrix as defined Eq.\ \eqref{td_form1}, the
values of $\sigma_i$, $i=1..k$ are given by the $k$ non-zero singular
values of the bidiagonal matrix
\begin{equation}\label{j_form}
J=\begin{pmatrix}
a_1&-b_1\\
&a_2&-b_2\\
&&\ddots&\ddots\\
&&&&a_{n-1}&-b_{n-1}\\
&&&&&a_n
\end{pmatrix}\,.
\end{equation}
For completeness, we give details and a self-contained derivation in
appendix \ref{canonical}.

The canonical form of a skew-symmetric matrix under unitary congruence
is also connected to certain eigenvalue problems: In the real case,
the eigenvalues of $A$ are given by $\pm i \sigma_j$. In the complex
case, the matrix $A^*A=-A^\dagger A$ has doubly degenerate eigenvalues
$\sigma_j^2$.

\subsection{Skew-symmetric tridiagonalization and existing approaches}
\label{others}

Both the computation of the Pfaffian and of the canonical form
are ultimately connected to the problem of tridiagonalizing a
skew-symmetric matrix. Here we give an overview of existing solutions
(with implementations) that could be used to solve parts of the problem,
and discuss the need for a new comprehensive implementation.

For real skew-symmetric matrices, the unitary congruence transformation
reduces to an ordinary orthogonal similarity transformation and hence
established decompositions can be used \cite{Golub1996}: The Hessenberg
decomposition of a skew-symmetric matrix reduces to tridiagonal form
\eqref{td_form1}, and the real Schur decomposition to the canonical
form \eqref{youla} (implemented, for example in LAPACK
\cite{Lapack}). However, none of these decompositions make use of the
structure of the problem which would be desirable for precision and speed,
nor can they be used for complex skew-symmetric matrices.

Ward and Gray have developed and implemented algorithms to compute the
tridiagonal form and the eigenvalues (and as an intermediate step,
the canonical form) of a real dense, skew-symmetric matrix,
making use of the structure of the problem
\cite{Ward1978b}. A complex version is however not available.

The accompanying Matlab code \cite{HighamMCT} to \cite{Higham2002} contains
a skew-symmetric $LDL^T$ decomposition that can be used to
compute Pfaffians, but according to the authors is not designed
for efficiency.

Very recently, Gonz\'alez-Ballestero, Robledo and Bertsch
have developed a library for the numerical computation of
the Pfaffian of a dense skew-symmetric
matrix \cite{Gonzalez-Ballestero2010}, but do not give access to the
transformation matrix (e.g. needed for computing the
canonical form). They present algorithms based on a $LDL^T$ decomposition
(called Aitken block diagonalization in \cite{Gonzalez-Ballestero2010})
and on Householder tridiagonalization. However, their approach does not
make use of the full symmetry of the problem.

None of the existing approaches (with the exception of LAPACK that
does not exploit the skew-symmetry of the problem) makes use
of block algorithms that are rich in level-3 operations and desirable
for a more favorable memory access pattern. Below we will show that
such block algorithms can give rise to a considerable increase in speed.

Moreover, none of the above approaches makes use of the
sparsity of a banded matrix, a structure that however often arises
in practice. Below we will also consider this case in particular.

The goal of this work is thus to develop and implement algorithms for
tridiagonalizing a real or complex skew-symmetric matrix, making use of
the skew-symmetry and possibly the bandedness of the matrix.

\subsection{Outline}

The remainder of the paper is organized as follows. In
Sec.\ \ref{tridi} we discuss algorithms to tridiagonalize a dense or
banded skew-symmetric matrix using Gauss transformations, Householder
reflections and Givens rotations.  Further, in Sec.\ \ref{impl} we
discuss the details of our implementation, and present benchmarks and
an exemplary application in Sec.\ \ref{examples}. In the appendix, we
give a self-contained derivation on the computation of the canonical
form of a tridiagonal, skew-symmetric matrix. Moreover, we discuss
blocked versions of our tridiagonalization algorithms for dense matrices
and give technical details about the Fortran implementation.

\section{Skew-symmetric numerical tridiagonalization}
\label{tridi}

\subsection{Statement of the problem}

Summarizing the discussion above, for a given skew-symmetric $n\times
n$ matrix $A$ we seek a (invertible) transformation $B$ such that $A=
B T B^T$ with $T$ in tridiagonal form tridiagonal (or in partial
tridiagonal for).  Below we consider first an algorithm for dense
matrices based on Gauss transformations requiring pivoting. Then we
focus on algorithms bases on unitary transformations where we consider
both dense and banded matrices.  The discussion is presented for the
case of complex matrices, but it carries over to the real case
unchanged.

\subsection{$LTL^T$ decomposition of dense matrices using the Parlett-Reid
algorithm}

For symmetric or Hermitian matrices there exist efficient algorithms
to compute a $LTL^T$ or $LDL^T$ decomposition (for an overview, see
\cite{Golub1996}). It has been shown by Bunch that those
decompositions can in principle also be generalized and computed
stably for skew-symmetric matrices \cite{Bunch1982}. Below we
reformulate the algorithm for the $LTL^T$ decomposition of a symmetric
matrix due to Parlett and Reid \cite{Parlett1970} such that it is
suitable for skew-symmetric matrices. The Parlett-Reid algorithm is usually
not the method of choice in the symmetric case, as there are more
efficient alternatives \cite{Aasen1971, Bunch1977}. However, as we will
discuss below, the Parlett-Reid algorithm can be used to compute the
\emph{Pfaffian} just as effective.

A $n\times n$ matrix of the form
\begin{equation}
M_k=E_n-\balpha_k (\be_k^{(n)})^T
\end{equation}
where $E_n$ is the $n\times n$ identity matrix and $\be_k^{(n)}$ the
$k$-th unit vector in $\mathbb{C}^n$, is called a \emph{Gauss
transformation} if the first $k$ entries of $\balpha_k$ are
zero. Given a vector $\bx=(x_1 \dots x_n)^T$ and taking
$\balpha_k=(0\dots 0\; x_{k+1}/x_k \dots x_n/x_k)^T$, $M_k$ can be used to
eliminate the entries $k+1 \dots n$ in $\bx$, $M_k \bx=(x_1 \dots x_k\;
0\dots 0)$, provided that $x_k\neq 0$.

A Gauss transformation can thus be used to zero the entries in a
column or row of $A$ below a chosen point $k$. In order to avoid
divisions by a small number or zero, a permutation $P_k$
interchanging entry $k$ with another nonzero, typically the largest
entry in $k+1 \dots n$ is performed. The numerical stability of
this pivoting strategy is discussed in \cite{Bunch1982}.

Hence, a series of Gauss transformations and permutations can be used
to tridiagonalize a skew-symmetric matrix $A$. To demonstrate the
mechanism, assume that after applying $k-1$ Gauss transformations and
permutations, the matrix $A^{(k-1)}=M_k P_k \dots M_2 P_2\, A\,
P_2^T M_2^T \dots P_k^T M_k^T$ is already in
tridiagonal form in the first $k-1$ columns and rows and hence has the
form
\begin{equation}\label{Akminus1}
A^{(k-1)}=\begin{pmatrix}
A_{11}&A_{12}&0\\
A_{21}&0&A_{23}\\
0&A_{32}&A_{33}
\end{pmatrix}
\begin{matrix}
k-1\\
1\\
n-k
\end{matrix}
\end{equation}
with $A_{11} \in \mathbb{C}^{k-1\times k-1}$, $A_{21} \in
\mathbb{C}^{1\times k-1}$, $A_{32} \in \mathbb{C}^{n-k\times 1}$,
$A_{33} \in \mathbb{C}^{n-k\times n-k}$, and $A_{12}=-A_{21}^T$,
$A_{23}=-A_{32}^T$ (transformations of the form $B A B^T$ maintain
skew-symmetry). Now choose a permutation matrix $\tilde{P}_{k+1}$ such
that the maximal entry in $A_{32}=( a_{k+1} \dots a_n)^T$ is permuted to
the top, i.e. $\tilde{P}_{k+1} A_{32}= (\tilde{a}_{k+1} \dots
\tilde{a}_n)^T$ where $\left |\tilde{a}_{k+1}\right|
=\max(\left|a_{k+1}\right|,\dots\left|a_n\right|)$. If the maximal
element at this step is zero, $A_{32}=0$ and $A^{(k-1)}$ is already tridiagonal
in the first $k$ columns and we set $M_{k+1}=P_{k+1}=E_n$. Otherwise,
we take $P_{k+1}=\mathrm{diag}(E_k,\tilde{P}_{k+1})$ and
$M_{k+1}=\mathrm{diag}(E_k,\tilde{M}_{k+1})$ with
$\tilde{M}_{k+1}= E_{n-k}-\balpha_{k+1} (\be_1^{(n-k)})^T$ with $\balpha_{k+1}=
(0\; \tilde{a}_{k+2}/\tilde{a}_{k+1} \dots \tilde{a}_n/\tilde{a}_{k+1})^T$.
Then we obtain
\begin{equation}
A^{(k)}=
M_{k+1} P_{k+1} A^{(k-1)} P_{k+1}^T M_{k+1}^T=\begin{pmatrix}
A_{11}&A_{12}&0\\
A_{21}&0&A_{23}\tilde{P}_{k+1}^T\tilde{M}_{k+1}^T\\
0&\tilde{M}_{k+1}\tilde{P}_{k+1}A_{32}&
\tilde{M}_{k+1}\tilde{P}_{k+1}A_{33}\tilde{P}_{k+1}^T\tilde{M}_{k+1}^T
\end{pmatrix}.
\end{equation}
Then $\tilde{M}_{k+1}\tilde{P}_{k+1}A_{32} \propto \be_1^{(n-k)}$
and $A^{(k)}$ is tridiagonal in its first $k$ rows and columns.
Defining $\bw=\tilde{P}_{k+1}A_{33}\tilde{P}_{k+1}^T \be_1^{(n-k)}$ we find
\begin{equation}\label{skew_update_ltl}
\tilde{M}_{k+1}\tilde{P}_{k+1}A_{33}\tilde{P}_{k+1}^T\tilde{M}_{k+1}^T
=\tilde{P}_{k+1}A_{33}\tilde{P}_{k+1}^T+\balpha_{k+1} \bw^T-\bw
\balpha_{k+1}^T\,.
\end{equation}
The cross-term
$(\be_1^{(n-k)})^T\tilde{P}_{k+1}A_{33}\tilde{P}_{k+1}^T
\be_1^{(n-k)}$ vanishes due to the skew-symmetry of $A_{33}$. Note
that in this skew-symmetric outer product update, the matrix
$\tilde{P}_{k+1}A_{33}\tilde{P}_{k+1}^T$
remains actually unchanged in the first column and row due to the
structure of $\balpha_{k+1}$ and $\bw$. The outer product update is
dominating the computational cost of each step and can be computed in
$2(n-k)^2$ flops, if the symmetry is fully accounted
for. Fig.~\ref{ltl_and_householder} shows the structure of a
tridiagonalization step schematically for a particular example.

\begin{figure}

\[
   \left(
    \begin{array}{cccccccc}
      0&\times\\\cline{3-6}
      \times&0&\multicolumn{1}{|c}{\times}&\boxdot&\boxdot&\multicolumn{1}{c|}{\boxdot}\\\cline{2-2}
      &\multicolumn{1}{|c}{\times}&0&\times&\times&\multicolumn{1}{c|}{\times}\\
      &\multicolumn{1}{|c}{\boxdot}&\times&0&\times&\multicolumn{1}{c|}{\times}\\
      &\multicolumn{1}{|c}{\boxdot}&\times&\times&0&\multicolumn{1}{c|}{\times}\\
      &\multicolumn{1}{|c}{\boxdot}&\times&\times&\times&\multicolumn{1}{c|}{0}\\\cline{2-6}

   \end{array}
   \right)
\]
\caption{Example of the structure induced by applying a Gauss or
  Householder transformation from left and right. $\times$ represents
  nonzero entries and $\boxdot$ those entries that are zeroed by the
  Gauss or Householder transformation.  In this example the first
  column and row have already been reduced, and the transformation is
  applied to $A^{(1)}$ from the left and the right.  The parts of the
  matrix that are changed in the process are marked with a frame.}
\label{ltl_and_householder}
\end{figure}

After $n-2$ steps, the decomposition can be written as
\begin{equation}
P A P^T = L T L^T
\end{equation}
with permutation $P=P_{n-1} \dots P_2$, skew-symmetric tridiagonal
$T=M_{n-1} P_{n-1} \dots M_2 P_2\, A\, P_2^T M_2^T \dots P_{n-1}^T
M_{n-1}^T$, and lower unit triangular matrix
\begin{equation}
L=(M_{n-1} P_{n-1} \dots M_2 P_2 P^T)^{-1}\,.
\end{equation}
As in the symmetric Parlett-Reid algorithm, the first column of $L$ is
$\be_1^{(n)}$, and the $k$-th column below the diagonal a permuted version of
$\balpha_k$.

The computation of the updated matrix, Eq.\ \eqref{skew_update_ltl},
is a level-2 matrix operation. It is possible to regroup these updates
in a way that allows to operate with level-3 matrix operations that
have a more favorable memory access pattern.  The details of this block
version of the Parlett-Reid algorithm are given in
appendix \ref{blockhouse}.

The full skew-symmetric $LTL^T$ decomposition can be computed in
$2 n^3/3$ flops. It is however readily generalized to compute only a
partial tridiagonal form as in Eq.\ \eqref{partialT} by skipping
every other row and column elimination. This partial $L T L^T$
decomposition can thus be computed in $n^3/3$ flops. Since
$\det(L)=1$ and $\det(P)$ can be computed in $n$ steps, computing the
Pfaffian of a skew-symmetric matrix with the Parlett-Reid algorithm thus
requires a total of $n^3/3$ flops. This is a factor of 10 less than the
unsymmetric Hessenberg decomposition.

For computing a full tridiagonalization, the Parlett-Reid algorithm
requires twice as many flops as other approaches: Aasen's algorithm
\cite{Aasen1971} computes a (complete) $LTL^T$ decomposition using a
different order of operations in $n^3/3$ flops, as does the
Bunch-Kaufmann algorithm \cite{Bunch1977} for computing a (complete) $LDL^T$
decomposition. Both algorithms can be generalized to the
skew-symmetric case.  Given the fact that computing the Pfaffian
requires less information than a full tridiagonalization, it might
seem feasible to compute the Pfaffian in $n^3/6$ flops. However,
neither Aasen's algorithm (which is based on the fact that $TL^T$ is
upper Hessenberg and hence $T$ fully tridiagonal), nor the
Bunch-Kaufmann algorithm (which relies on the block-diagonal structure
of $D$) are easily amended to compute a suitable partial
factorization. Thus, for computing the Pfaffian, the Parlett-Reid
algorithm is competitive. It remains an open question if it is
possible to compute the Pfaffian of a dense skew-symmetric matrix in
less than $n^3/3$ flops.

\subsection{Tridiagonalization of dense matrices with Householder reflections}

Dense symmetric or Hermitian matrices are commonly reduced to
tridiagonal form by Householder transformations \cite{Golub1996},
and we adopt this approach to the skew-symmetric case here.

An order $m$ \emph{Householder} transformation $H$ is a matrix of the form
\[
H=1-\tau \bv \bv^\dagger
\]
where $\tau \in \mathbb{C}$ and $\bv \in \mathbb{C}^m$ chosen such that
\[
H \bx = \alpha ||x||_2 \be_1^{(m)}
\]
for a given $\bx \in \mathbb{C}^m$. Here $||\cdot||_2$ denotes the norm of a
vector, $\be^{(m)}_1$ is the first unit vector in $\mathbb{C}^m$ and
$\alpha \in \mathbb{C}$. For example, $\alpha=-e^{i\phi}$, if one chooses
$\bv=\bx+e^{i\phi}||x||_2 \be_1^{(m)}$ when $x_1=e^{i\phi} |x_1|$,
but there is a certain degree of freedom in choosing the Householder
vector $\bv$ which can be exploited to maximize
stability (for an overview, see \cite{lawn72}).
Note that $H$ is unitary (though not necessarily Hermitian) and can also
be numerically calculated such that it is unitary up to machine precision
\cite{Golub1996}.

Householder transformations can thus be applied to a matrix to zero
all the elements of a column (or row) below a chosen point, just as
Gauss transformations, but without the need for pivoting. As a
consequence, the structure of the tridiagonalization procedure is
analogous to the $LTL^T$ decomposition. Assume that after step $k-1$
the matrix $H_{k-1}\dots H_1 A H_1^T \dots H_{k-1}^T$ is already
tridiagonal in the first $k-1$ columns and rows and partitioned as
defined in Eq.\ \eqref{Akminus1}.  Then an order $n-k$ Householder
matrix $\tilde{H}_k$ is chosen such that $\tilde{H}_k A_{32} \propto
\be_1^{(n-k)}$ and the full transformation is set to
$H_k=\mathrm{diag}(E_k, \tilde{H}_k)$.  Writing $\tilde{H}_k=1-\tau
\bv \bv^\dagger$ and defining $\bw=\tau A_{33} \bv^*$ we find
\begin{equation}\label{skew_update_hh}
\tilde{H}_kA_{33}\tilde{H}_k^T=A_{33}+\bv \bw^T-\bw \bv^T\,.
\end{equation}
The main difference to the $LTL^T$ decomposition is the fact that
the computation of $\bw$ now involves a full matrix-vector
multiplication. Hence, the total computational cost of the
outer product update in Eq.\ \eqref{skew_update_hh} is
$4(n-k)^2$ flops.
The structure of a Householder tridiagonalization
step is also shown schematically in Fig.~\ref{ltl_and_householder}.
The outer product updates of Eq.\ \eqref{skew_update_hh} can be rearranged
to increase the fraction of level-3 matrix operations. The block
version of the Householder algorithm is detailed in App.\ \ref{blockhouse}.

Complete tridiagonalization with Householder matrices requires in
total $4n^3/3$ flops. This can reduced to $2n^3/3$ for computing the
Pfaffian by skipping every other row/column elimination to compute
only a partial tridiagonal form.

For the computation of the Pfaffian we also need to compute the
determinant of the transformation matrix $Q=H_1^\dagger H_2^\dagger \dots
H_{n-2}^\dagger$. The determinant of a single Householder transformation
$H^\dagger=1-\tau^* \bv \bv^\dagger$ is given as
\begin{equation}
\det(H^\dagger)=1-\tau^* \bv^\dagger \bv\,.
\end{equation}
For the particular choice $\tau=2/\bv^\dagger \bv$,
$\det(H)=\det(H^\dagger)=-1$, i.e. $P$ is a reflection, but other
choices of $\tau$ are equally viable. In particular, if the matrix is
already tridiagonal in certain column and row (which can happen
frequently for very structured matrices), it is advantageous to use
$H=E_n$.  Moreover, any complex skew-symmetric matrix may be reduced to
a purely real tridiagonal matrix using appropriate Householder
transformations with complex $\tau$ \cite{lawn72}. Because of this, the
determinant of each Householder reflection must be computed
separately.  The task of computing $\det(Q)$ still only scales as
$\propto n^2$ and is thus negligible compared to the
tridiagonalization cost.

In summary, for computing the Pfaffian Householder tridiagonalization
is twice as costly as the Parlett-Reid algorithm and thus usually not
competitive. It has however a right on its own given its connection
to computing the canonical form of a skew-symmetric matrix.

\subsection{Tridiagonalization of band matrices with Givens rotations}
\label{givens}

The dense algorithms of the previous two sections are not easily
amended to matrices with a finite band width. In the case of the
Parlett-Reid algorithm, the symmetric pivoting can lead to an uncontrolled
growth of the band width depending on the details of the matrix.
In the Householder tridiagonalization, the outer product matrix
update always introduces values outside the band, leading to a
fast-growing band width.

For symmetric matrices, $LTL^T$ or $LDL^T$ decomposition algorithms
respecting the band width have only been introduced recently
\cite{Irony2006,Kaufman2007}. In contrast, banded tridiagonalization
with unitary transformations is well established for symmetric matrices,
and we adopt this approach for the skew-symmetric case below.

Instead of zeroing a whole column or row as is done in the Householder
approach, for banded matrices it is of advantage to use a more selective
approach. The method of choice for this case in the symmetric or Hermitian case
are Givens rotations \cite{Schwarz1968}, and we will extend this
approach to th skew-symmetric case. A \emph{Givens} rotation $G_{i,j}$
is a modification of the identity matrix that is only different in the
$i$th and $j$th row and column. It is defined as
\begin{equation}
G_{i,j}=\begin{pmatrix}
1&\cdots&0&\cdots&0&\cdots&0\\
\vdots&\ddots&\vdots&&\vdots&&\vdots\\
0&\cdots&c&\cdots&s&\cdots&0\\
\vdots&&\vdots&\ddots&\vdots&&\vdots\\
0&\cdots&-s^*&\cdots&c&\cdots&0\\
\vdots&&\vdots&&\vdots&\ddots&\vdots\\
0&\cdots&0&\cdots&0&\cdots&1
\end{pmatrix}
\begin{matrix}
\\
\\
i\\
\\
j\\
\\
\\
\end{matrix}
\end{equation}
with $c\in\mathbb{R}$, $s\in\mathbb{C}$ and $c^2+|s|^2=1$ and thus clearly
unitary.
Choosing $c$ and $s$ such that
\begin{equation}
\begin{pmatrix}
c&s\\
-s^*&c\\
\end{pmatrix}
\begin{pmatrix}
x_i\\
x_j
\end{pmatrix}
=\begin{pmatrix}
\tilde{x}_i\\
0
\end{pmatrix}
\end{equation}
it is possible to selectively zero one element of a vector.
Again, a Givens rotation can be computed numerically such that it is
orthogonal up to machine precision.

\begin{figure}
\[\left(
    \begin{array}{cccccccccc}\cline{3-4}
    0&\times&\multicolumn{1}{|c}{\times}&\multicolumn{1}{c|}{\boxdot}\\
   \times&0&\multicolumn{1}{|c}{\times}&\multicolumn{1}{c|}{\times}&\times\\\cline{1-8}
    \multicolumn{1}{|c}{\times}& \times&\multicolumn{1}{|c}{0}&\multicolumn{1}{c|}{\times}&\times&\times&\boxtimes\\
    \multicolumn{1}{|c}{\boxdot}&\times&\multicolumn{1}{|c}{\times}&\multicolumn{1}{c|}{0}&\times&\times&\times\\ \cline{1-8}
  &\times&\multicolumn{1}{|c}{\times}&\multicolumn{1}{c|}{\times}&0&\times&\times&\times\\
    &&\multicolumn{1}{|c}{\times}&\multicolumn{1}{c|}{\times}&\times&0&\times&\times\\
    &&\multicolumn{1}{|c}{\boxtimes}&\multicolumn{1}{c|}{\times}&\times&\times&0&\times\\
     &&\multicolumn{1}{|c}{} &\multicolumn{1}{c|}{}&\times&\times&\times&0\\
      &&&&&&&&\ddots
   \end{array}
   \right)
\]
\caption{Example for the structure induced by applying a Givens rotation $G_{2,3}$
to a skew-symmetric, banded matrix from the left and right: $\times$ denotes nonzero entries,
$\boxdot$ the entry that is eliminated by the Givens transformation, and
$\boxtimes$ the entries that are introduced outside band (fill-in).
The Givens rotation only affect the second and third row and column
(marked by frames).}\label{givens_struct}
\end{figure}

A banded skew-symmetric matrix can be brought into tridiagonal form
by Givens rotations of the form $G_{i,i+1}$. The structure induced in the
process of applying $G_{i,i+1}$ from the left and right is shown
schematically in fig.~\ref{givens_struct}. Applying a Givens rotation
$G_{i,i+1}$  ($G_{i,i+1}^T$) from the left (right) only modifies the
$i$th and $(i+1)$th rows (columns). Due to the skew-symmetry, if
$G_{i,i+1}$ zeroes the $(i+1)$ entry in column $j$, $G_{i,i+1}^T$
zeroes the $(i+1)$ entry in row $j$. Furthermore, each
Givens rotation only introduces at most one additional
nonzero entry outside the band in a row and column $k>i+1$.
This nonzero entry can thus be moved further down
the band by a sequence of Givens transformations until it is
``chased'' beyond the end of the matrix.

The structure of the skew-symmetric tridiagonalization routine is thus
identical to the symmetric or Hermitian case. The main difference is
in the update of the diagonal $2\times 2$-block that is affected
by both Givens rotations from left and right: Due to the skew-symmetry, the
diagonal blocks are invariant under these transformations,
\begin{equation}
\begin{pmatrix}
c&s\\
-s^*&c
\end{pmatrix}
\begin{pmatrix}
0&a\\
-a&0
\end{pmatrix}
\begin{pmatrix}
c&-s*\\
s&c
\end{pmatrix}
=\begin{pmatrix}
0&a\\
-a&0
\end{pmatrix}.
\end{equation}
Kaufman \cite{Kaufman1984,Kaufman2000} has presented a variant of the
symmetric band matrix approach of Ref.\ \cite{Schwarz1968} that allows
to work on more data in a single operation, which allows a more
favorable memory access pattern.  These modifications carry over
unchanged to the skew-symmetric case.

The tridiagonalization of an $n\times n$ skew-symmetric matrix
with bandwidth $b$ using Givens transformations scales as
$\mathcal{O}(b n^2)$.

The determinant of any single Givens rotation $\det(G_{i,j})=1$, and
thus the determinant of the full transformation $\det(Q)=1$, too. In
the complex case the resulting tridiagonal matrix can be chosen to be
purely real, in this case the determinant of total unitary
transformations (the Givens transformations and row/columns-scalings
with a phase factor) obey $|\det(Q)|=1$.

\section{Notes on the implementation}
\label{impl}

\subsection{Fortran}

We have implemented the algorithms described in this manuscript as a
comprehensive set of Fortran routines for real and complex variables
as well as single and double precision. Because of the conceptional
similarity of the skew-symmetric problem to the symmetric and
Hermitian problem, these routines are designed analogous to to the
corresponding symmetric and Hermitian counterparts in
LAPACK. Moreover, our implementation also makes use of the LAPACK
framework for computing, applying, and accumulating Householder and
Givens transformations, which was designed for numerical stability and
which is available in an optimized form for any common computer
architecture.

Dense skew-symmetric matrices are stored as ordinary two-dimensional
Fortran matrices, but only the strictly lower or upper triangle needs
to be set (for differences in the implementation between lower and
upper triangular storage see App.\ \ref{upper_vs_lower}). For banded
skew-symmetric matrices, only the strictly upper or lower diagonals
are stored in a $\mathtt{K} \times \mathtt{N}$ array $\mathtt{AB}$,
where $\mathtt{K}$ is the number of non-zero off-diagonals and
$\mathtt{N}$ the size of the matrix. The $\mathtt{j}$-th column of the
matrix $A$ is stored in the $\mathtt{j}$-th column of $\mathtt{AB}$ as
\begin{itemize}
\item $\mathtt{AB(K+1+i-j,j)} = A_{i,j}$ for
  $\max(1,\mathtt{j-K})<=\mathtt{i}<=\mathtt{j}$, if the upper triangle
  is stored,
\item $\mathtt{AB(1+i-j,j)} = A_{i,j}$ for $
  \mathtt{j}<=\mathtt{i}<=\min(\mathtt{N,j+kd})$, if the lower triangle is
  stored.
\end{itemize}
Note that in this scheme, also the zero diagonal is explicitly stored.
This was done to keep the design identical to the storage scheme of
symmetric and Hermitian band matrices in LAPACK.

\begin{table}
\begin{tabular}{|l|p{0.65\linewidth}|}
\hline \texttt{SKTRF}&Skew-symmetric tridiagonal decomposition of a dense
matrix using the block Parlett-Reid algorithm.\\
\hline \texttt{SKTF2}&Skew-symmetric tridiagonal decomposition of a dense matrix
using the Parlett-Reid algorithm (unblocked version).\\
\hline \texttt{SKTRD}&Skew-symmetric tridiagonalization of a dense matrix
using block Householder reflections.\\
\hline \texttt{SKTD2}&Skew-symmetric tridiagonalization of a dense matrix
using Householder reflections (unblocked version).\\
\hline \texttt{SKPFA}&Compute the Pfaffian of a dense skew-symmetric matrix
(makes use of either \texttt{SKTRD} or \texttt{SKTRF}).\\
\hline \texttt{SKPF10}&Compute the Pfaffian of a dense skew-symmetric matrix
(makes use of either \texttt{SKTRD} or \texttt{SKTRF}). The result is returned
as $a \times 10^b$ to avoid over- or underflow.\\
\hline \hline \texttt{SKBTRD}&Skew-symmetric tridiagonalization of a
banded matrix using Givens rotations.\\
\hline \texttt{SKBPFA}&Compute the Pfaffian of a banded skew-symmetric matrix
(makes use of \texttt{SKBTRD}).\\
\hline \texttt{SKPF10}&Compute the Pfaffian of a banded skew-symmetric matrix
(makes use of \texttt{SKBTRD}).The result is returned as $a \times 10^b$
to avoid over- or underflow.\\
\hline
\end{tabular}
\caption{Overview of the computational routines in the Fortran
  implementation.  In the Fortran77 interface the routine name must be
  preceded by either \texttt{S} (single precision), \texttt{D} (double
  precision), \texttt{C} (complex single precision), or \texttt{Z}
  (complex double precision).}
\label{overview}
\end{table}

Our library includes stand-alone routines for the tridiagonalization
of a skew-symmetric dense matrix (\texttt{SKTRF} and \texttt{SKTF2}
using the Parlett-Reid algorithm, \texttt{SKTRD} and \texttt{SKTD2}
using the Householder approach) and banded matrices (\texttt{SKBTRD}).
We also include functions to compute the Pfaffian of a skew-symmetric
dense (\texttt{SKPFA} and \texttt{SKPF10}) and banded matrices
(\texttt{SKBPFA} and \texttt{SKBPF10}), which are based on the
tridiagonalization functions.  As the determinant, the Pfaffian of a
large matrix is prone to floating point over- or underflow. Because of
that, we have included routines that return the Pfaffian in the form
$a\times 10^b$, where $a$ is real or complex, and $b$ is always real
and integer (\texttt{SKPFA10} and \texttt{SKBPF10}). Both a Fortran95
and a Fortran77 interface are provided. In the Fortran77 version of
the code the routine name is preceded by either \texttt{S} (single
precision), \texttt{D} (double precision), \texttt{C} (complex single
precision), or \texttt{Z} (complex double precision).  The
computational routines and their purpose are summarized in
Table\ \ref{overview}.

The block versions of the algorithm have an internal parameter
controlling the block size. By default, the routines use the same
block sizes as their symmetric counterpart from the LAPACK library.
However, this internal parameter may be changed by the user to
optimize for a specific architecture.

Apart from the documentation here, all routines (including the
auxiliary ones) are documented extensively in the respective files.
Due to our routines using LAPACK and BLAS, both libraries must be
also linked.

\subsection{Python, Matlab and Mathematica}

While most compiled languages (including \texttt{C} and \texttt{C++})
are easily interfaced with a Fortran library, interpreted languages such
as \texttt{Python} or programs such as Matlab or Mathematica require somewhat more
effort. For this reason we have included stand-alone versions of the
tridiagonalization of dense skew-symmetric matrices using Householder
reflections implemented in \texttt{Python}, Matlab and Mathematica. Those
implementations, being of course slower than the Fortran counterpart,
are useful especially for situations where speed is not critical.
Both implementations also make use of the fact that for computing the
Pfaffian, only the odd rows and columns need to be tridiagonalized,
but always work on the full matrix instead of a single triangle.

Again, more extensive documentation for both implementations may be found
in the respective files.

\section{Examples}
\label{examples}

\subsection{Benchmarks}

To demonstrate the effectiveness of our methods, we have performed benchmark
computations of the Pfaffian of large, random matrices on various architectures. In
Table\ \ref{benchmark} we compare our approach with the other available
software that can also be used to calculate the Pfaffian in certain
situations (see Sec.\ \ref{others}). For this benchmark we have
compiled our Fortran implementation, and the implementations of
Refs.\ \cite{Ward1978b} and \cite{Gonzalez-Ballestero2010} using
the same compiler and compilation options, and chose a machine-optimized
version of the LAPACK library \cite{Lapack}.

\begin{table}
\small
\begin{tabular}{|>{\centering}m{0.17\linewidth}|*{9}{|>{\centering}m{0.082\linewidth}}|}
\hline
&Block Parlett-Reid&Regular Parlett-Reid&
Block House\-holder&Regular House\-holder&Givens for band matrix&
\texttt{DGEHRD} (Lapack) \cite{Lapack}&\texttt{TRIZD} from \cite{Ward1978b}&
\texttt{PfaffianH} from \cite{Gonzalez-Ballestero2010}&
\texttt{PfaffianF} from \cite{Gonzalez-Ballestero2010}\tabularnewline
\hline\hline
\multicolumn{10}{|l|}{(a) benchmark for AMD Opteron 6174 2.2 Ghz}\tabularnewline
\hline
real $3000\times 3000$,
dense&5.1&5.9&9.4&10.5&-&24.7&383.4&54.5&120.8\tabularnewline
\hline
real $3000\times 3000$,
banded, $k=100$&0.9&0.7&9.3&10.5&2.1&24.8&383.2&54.3&121.7\tabularnewline
\hline
complex $2000\times 2000$,
dense&3.5&4.4&7.6&8.2&-&-&-&32.2&50.1\tabularnewline
\hline
complex $2000\times 2000$,
banded $k=100$&0.8&0.7&7.6&8.1&2.0&-&-&31.9&50.2\tabularnewline
\hline\hline
\multicolumn{10}{|l|}{(b) benchmark for Intel Core 2 Duo E8135 2.66 Ghz}\tabularnewline
\hline
real $3000\times 3000$,
dense&3.5&8.3&8.4&12.4&-&30.7&105.3&76.3&49.4\tabularnewline
\hline
real $3000\times 3000$,
banded, $k=100$&0.7&0.5&8.4&12.2&1.4&30.4&105.4&75.9&48.5\tabularnewline
\hline
complex $2000\times 2000$,
dense&3.5&5.0&7.5&8.3&-&-&-&48.3&28.3\tabularnewline
\hline
complex $2000\times 2000$,
banded, $k=100$&0.8&0.7&7.5&8.2&3.0&-&-&49.3&26.3\tabularnewline
\hline
\end{tabular}
\caption{Benchmark comparison of the implementation of this work and
other numerical approaches to compute the Pfaffian of a skew-symmetric matrix.
The table shows the time needed to compute the Pfaffian for the various methods
(time given in seconds) on two different architectures
[(a) and (b)]. The first five columns of benchmark results correspond
to the methods discussed in this work. For the banded matrices,
$k$ denotes the strictly upper or lower bandwidth, the full bandwidth
is hence $2k+1$.}\label{benchmark}
\end{table}

From the benchmark results we can see that the block
approach is always faster than the unblocked version. The relative
speed-up depends strongly on the architecture, but can reach up to
60\%. We also observe that the relative speed-up of the Parlett-Reid
algorithm is larger than of the Householder tridiagonalization,
reflecting the fact that the level-3 content of the former is larger
(see App.~\ref{blockhouse}).

For the banded random matrices we observe that the Parlett-Reid algorithm
performs surprisingly well. Although it is not designed to make use of the
bandedness of the matrix, the implementation of the skew-symmetric
outer product update takes into account zeros in the vectors of
the update. The Householder tridiagonalization does not benefit as
much, as for the matrices here the band width growth in the Householder
approach is faster than in the Parlett-Reid algorithm. It should be
stressed however that the performance of these algorithms in the banded
case depends on the actual values of the matrix. For example, band width
growth is stronger in the Parlett-Reid algorithm, if the largest
entries sit at the edge of the band.

The specialized approach for banded matrices using Givens
transformations is still slightly slower than the Parlett-Reid
algorithm for the matrix sizes considered here. The main benefit of
the specialized algorithm for Pfaffian calculations is hence its much
lower memory requirement. In fact, typically memory limits the matrix
sizes that can be handled, not computational time. The banded
Givens-based approach however is considerably faster than the Householder
tridiagonalization which makes it very attractive for computing the
canonical form (or eigenvalues in the real case).

Comparing to other approaches, we observe that our routines are always
faster, typically by a factor of about 10 or more.  In particular in the
real case, our specialized approach is considerably faster than using
the real Hessenberg reduction, although we do not always reach the
full speed-up of a factor of 10 that can be expected from the
operation count, which is due to the optimization of the LAPACK
library used. The implementation for real matrices of Ref.\
\cite{Ward1978b} is particularly slow as its memory access pattern is
somewhat unfavorable for modern computer architectures.

\subsection{Application: topological charge of a disordered nanowire}

Finally, we apply our approach to computing the Pfaffian to a physical
example, namely the numerical computation of the topological charge of
a disordered nanowire.

A nanowire made out of a topological superconductor supports at its
two ends Andreev bound states pinned at the Fermi energy
\cite{Kitaev2001,Wimmer2010,Lutchyn2010,Oreg2010,Potter2010}.
Because of particle-hole symmetry, those states are Majorana fermions
-- particles that are their own anti-particle -- and may allow for
topologically protected quantum computing \cite{Kitaev2001}. In
contrast, an ordinary (trivial) superconducting wire does not support
such states. The recent proposal to realize a topological superconductor
using ordinary semiconducting and superconducting materials
\cite{Lutchyn2010,Oreg2010,Potter2010}
has stirred a lot of interest towards Majorana physics in condensed matter.

A topological charge $Q$ is a quantity that indicates whether a system is
in the trivial or topological state, and hence signifies the absence
or existence of Majorana bound states. A superconducting system
exhibits particle-hole symmetry which allows the Hamiltonian to be
written in the form \cite{Kitaev2001}
\begin{equation}\label{H_skew}
H=\frac{i}{4} \sum_{i,j} A_{ij} c_{i} c_{j}
\end{equation}
where $A$ is a skew-symmetric matrix, and $c_i$, $c_j$ Majorana
operators with $c_i^\dagger=c_i$, $c_i^2=1$ and
$c_ic_j+c_jc_i=\delta_{ij}$. Below we further specialize to
the case where particle-hole symmetry is the only remaining
symmetry, i.e.\ broken time-reversal and spin-rotation
symmetry, which puts the system into class D of the general
symmetry classification scheme \cite{Evers2008}.

Kitaev has shown that for a translationally invariant wire,
\begin{equation}\label{kitaev_Q}
Q=\sign\, \Pf[A(0)] \Pf[A(\pi/a_{\rm uc})]
\end{equation}
is a topological charge that signifies the absence ($Q=1$) or existence
($Q=-1$) of Majorana bound states at the ends. In this expression,
$A(k)$ is the Hamiltonian in (Bloch) momentum space written in the
Majorana basis. $A(k)$ is a matrix with a size corresponding to the
size of the unit cell, the unit cell length is denoted by $a_{\rm uc}$.
Note that the Pfaffian needs only be evaluated at two
values of momenta which correspond to closing the unit cell with
periodic ($k=0$) and anti-periodic ($k=\pi/a_{\rm uc}$)
boundary conditions.

For a clean system, the size of $A(k)$ is $\propto W$, where $W$
is the width of the wire, and Eq.\ \eqref{kitaev_Q} has been
used previously to compute the topological charge
\cite{Kitaev2001,Lutchyn2010}. A disordered system can be considered
(up to finite size effects, see below) in Eq.\ \eqref{kitaev_Q}
as a large, disordered supercell repeated periodically. In this case,
the size of $A(k)$ is $\propto W L$, where $L$ is the length of the
supercell. This implies that $A(k)$ in a disordered system will be
a very large matrix, and we are not aware of any application of
Eq.\ \eqref{kitaev_Q} for such as system. However, the
sparse structure of $A$, in particular its bandedness, allows the
application of the special algorithms developed in this work, and
allows for the first time applying \eqref{kitaev_Q} to large,
disordered systems.

Recently, an alternative definition of the topological charge for
class D systems has been shown \cite{Akhmerov2011}. In contrast
to Eq.\ \eqref{kitaev_Q} which is based on properties of the Hamiltonian,
this alternative definition is based on transport properties:
\begin{equation}\label{our_Q}
Q=\sign \det r\,,
\end{equation}
where $r$ is the reflection matrix. This definition is equally
applicable to clean and disordered systems. Below we show numerically
the equivalence of the definitions \eqref{kitaev_Q} and \eqref{our_Q}.

For this we use the model of Refs.\ \cite{Lutchyn2010,Oreg2010}
with a normal state Rashba Hamiltonian
\begin{equation}
H_0=\frac{\bm{p}^{2}}{2m_{\rm eff}}+U(\bm{r})+
\frac{\alpha_{\rm so}}{\hbar}(\sigma_{x}p_{y}-\sigma_{y}p_{x})+\tfrac{1}{2}g_{\rm eff}\mu_{B}B\sigma_{x}\,,
\end{equation}
where $m_\mathrm{eff}$ is the effective mass of the
two-dimensional electron gas, $\alpha$ the Rashba spin-orbit coupling,
and $g_{\rm eff}\mu_{B}B$ the Zeeman splitting due to an external
magnetic field.  Characteristic length and energy scales are $l_{\rm
so}=\hbar^{2}/m_{\rm eff}\alpha_{\rm so}$ and $E_{\rm so}=m_{\rm
eff}\alpha_{\rm so}^{2}/\hbar^{2}$.  The electrons are then confined
into a nanowire of width $W$ and length $L$ in the $x-y$-plane. For
the numerical treatment, the Hamiltonian is discretized on a square
grid with lattice constant $a$ and thus represented by a matrix
$H_{ij,\mu\nu}$, where $i$, $j$ denote lattice sites, and $\mu$, $\nu$
the spin degrees of freedom. Disorder is introduced as
a random on-site potential taken from the uniform distribution
$[-U_0,U_0]$. The Hamiltonian of the system in contact
with a s-wave superconductor then reads
\begin{equation}\label{H_sc}
H=\sum_{i,j,\mu,\nu} H_{ij,\mu\nu} a^\dagger_{i,\sigma} a_{j,\nu}
+\sum_i \Delta a^\dagger_{i,\uparrow}a^\dagger_{i,\downarrow}
+\Delta a_{i,\downarrow}a_{i,\uparrow}\,,
\end{equation}
where $\Delta$ is the proximity-induced pair potential. Defining Majorana
operators as $c_{2i-1,\mu}=\tfrac{1}{\sqrt{2}}(a_{i,\mu}+a_{i,\mu}^\dagger)$ and
$c_{2i,\mu}=\tfrac{i}{\sqrt{2}}(a_{i,\mu}-a_{i,\mu}^\dagger)$ we can transform
Eq.\ \eqref{H_sc} into the form \eqref{H_skew} with a skew-symmetric
matrix $A$ (whose bandwidth is reduced for the numerics with a bandwidth
reduction algorithm \cite{Gibbs1976}).

\begin{figure}
\begin{center}
\includegraphics[width=0.6\linewidth]{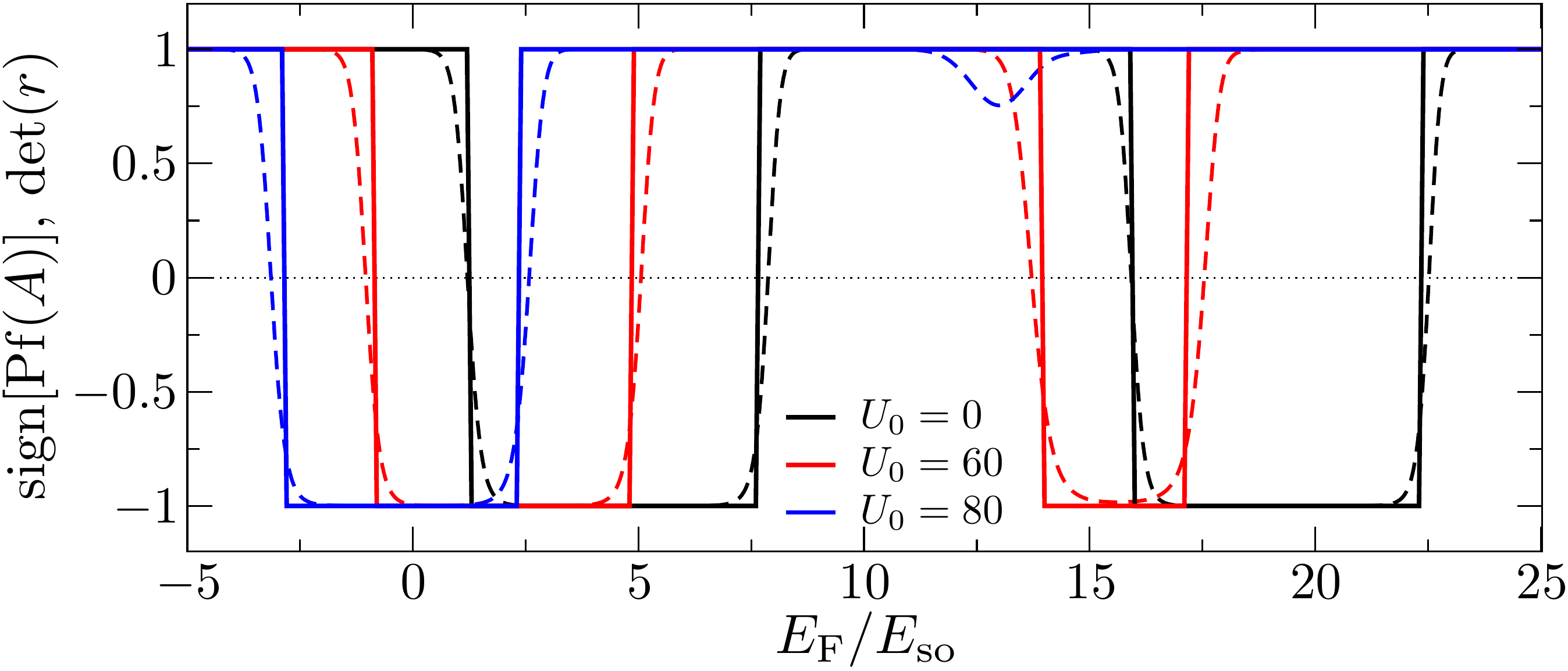}
\end{center}
\caption{Topological charge of a semiconductor nanowire in proximity
  to a superconductor for various disorder strengths as a function of
  the Fermi energy, computed from the Pfaffian of the Hamiltonian in
  the Majorana basis \eqref{kitaev_Q} (solid lines) and from the
  determinant of the reflection matrix \eqref{our_Q} (dashed lines).
  Parameters of the calculation were $W=l_{\rm so}$, $L=10 l_{\rm
  so}$, $a=l_{\rm so}/20$, $\Delta=10 E_{\rm so}$, and $g_{\rm
  eff}\mu_{B}B=21$.}
\label{results_Q}
\end{figure}

In Fig.\ \ref{results_Q} we show the numerical results for the topological
charge as defined in Eqs.\ \eqref{kitaev_Q} and \eqref{our_Q}. For the
computation of the Pfaffian in \eqref{kitaev_Q} we apply the Givens based
method from Sec.\ \ref{givens}, for computing the reflection matrix
the numerical method of Ref.\ \cite{Wimmer2009}.

In all cases, clean and disordered, both definitions of $Q$ agree very
well.  In particular, both definitions predict a vanishing of the
topological phase for the largest disorder in the region
$10<E_\text{F}/E_\text{so}<25$. There are only very small differences in
the exact value of $E_\text{F}$ where the topological charges change sign.
These differences can be explained by finite size effects: At these points
the bulk of the nanowire has a significant conductance ($|\det r|\ll 1$)
which in turn means that the different geometry of Eqs.\ \eqref{kitaev_Q}
(infinite repetition of a supercell) and \eqref{our_Q} (single
supercell connected to metallic leads) matter.
In contrast, in the regime where the bulk of the wire is fully insulating
($|\det r| \approx 1$), both definitions agree fully.

The algorithmic developments in this work have allowed to evaluate
Eq.\ \eqref{kitaev_Q} for a fairly large disordered system. The bandwidth
of the respective skew-symmetric matrix $A$ scales $\propto W$, and hence the
cost of tridiagonalization as $\propto W^3 L^2$. In contrast, the
definition of the topological charge from transport properties
\eqref{our_Q} scales $\propto W^3 L$ \cite{Wimmer2009}. Hence, from
a computational viewpoint, Eq.\ \eqref{our_Q} is more favorable. It is thus
reassuring that our numerical experiments showed the equivalence of both
definitions.

\section{Conclusions}

We have shown that both the computation of the Pfaffian and the canonical
form of a skew-symmetric matrix can be solved easily once the matrix
is reduced to skew-symmetric (partial) tridiagonal form. To find this form, we have
then developed tridiagonalization algorithms based on Gauss transformations,
using a skew-symmetric, blocked version of the Parlett-Reid algorithm, and
based on unitary transformations, using block Householder
reflections and Givens rotations, applicable to dense and skew-symmetric
matrices, respectively. These algorithms have been implemented
in a comprehensive numerical library, and its performance has been
proven to be superior to other approaches in benchmark calculations.
Finally, we have applied our numerical method for computing the Pfaffian
to the problem of computing the topological charge of a disordered
nanowire, showing numerically the equivalence of different methods based on
the Hamiltonian or the scattering matrix of the system.

\begin{acknowledgments}
  MW acknowledges stimulating discussions with F. Hassler, I.C. Fulga,
  A.R. Akhmerov, C.W. Groth and C.W.J. Beenakker, as well as support
  from the German academic exchange service DAAD.
\end{acknowledgments}

\appendix

\section{The computation of the canonical form of a skew-symmetric matrix}\label{canonical}

The problem of computing the canonical form of an even-dimensional
skew-symmetric matrix has been discussed in
\cite{Ward1978a,Golub1996,Hastings2010} (the odd-dimensional problem can
be reduced to an even-dimensional one by a series of Givens rotations
\cite{Ward1978a}). Here we give a self-contained derivation for completeness.

A $2n\times 2n$ skew-symmetric matrix $A$ can be reduced to tridiagonal form
$A=Q T Q^T$ with $Q$ unitary (orthogonal in the real case) and $T$ tridiagonal
as given in Eq.\ \eqref{td_form1}. This tridiagonal matrix can be rewritten as
\begin{equation}
T=P \begin{pmatrix}&J^T\\-J&\end{pmatrix} P^T
\end{equation}
with $J$ as given in Eq.\ \eqref{j_form} and $P$ the permutation
\begin{equation}
P=\begin{pmatrix}
1&2&\dots&n&n+1&n+2&\dots&2n\\
1&3&\dots&2n-1&2&4&\dots&2n
\end{pmatrix}\,.
\end{equation}
From the singular value decomposition (SVD) of $J=V \Sigma W$, with
$\Sigma=\text{diag}(\sigma_1,\dots,\sigma_k,0,\dots,0)$, $\sigma_i>0$,
$k=\text{rank}(J)$,
and $U$, $V$ unitary (orthogonal in the real case) matrices, we then obtain
\begin{equation}
T=P\begin{pmatrix}W^T&\\&V\end{pmatrix}
\begin{pmatrix}&\Sigma\\-\Sigma&\end{pmatrix}
\begin{pmatrix}W&\\&V^T\end{pmatrix}P^T\,.
\end{equation}
But since $P^T P=1$ and $P\begin{pmatrix}&\Sigma\\-\Sigma&\end{pmatrix}P^T=\Xi$ with
$\Xi$ defined in Eq.\ \eqref{youla}, we find the canonical form of
$A$ as
\begin{equation}
A=U \Xi U^T
\end{equation}
with
\begin{equation}
U=Q P \begin{pmatrix}W^T&\\&V\end{pmatrix} P^T\,.
\end{equation}

In practice, it suffices to implement the SVD for
real $J$, as any complex matrix can be reduced to real tridiagonal form
using unitary transformations. For the computation of the SVD one can make
use of the computational routines for bidiagonal matrices from LAPACK
\cite{Lapack}.

\section{Block versions of the Parlett-Reid and Householder tridiagonalization algorithms}
\label{blockhouse}

The application of the Gauss and Householder transformations in the
update operations Eqs.\ \eqref{skew_update_ltl} and
\eqref{skew_update_hh} are inherently level-2 matrix operations. It is
however possible to accumulate transformations and apply those
simultaneously in a block representations \cite{Golub1996}
which has a higher level-3 fraction. This procedure is also used for
the tridiagonalization of symmetric matrices \cite{lawn2} and we
describe its application to the skew-symmetric case below.

Both algorithms are based on transformations of the form
\begin{equation}
E_n - \bv_k \bu_k^T\,,
\end{equation}
and the update operation is of the form
\begin{equation}
A^{(k)} = A^{(k-1)}+\bv_k \bw_k^T -\bw_k \bv_k^T\,,
\end{equation}
with $\bw_k=A^{(k-1)} \bu_k$.

Assume that the matrix after the $k$-th update is given as
\begin{equation}\label{blockhouseeq}
A^{(k)}=A+V_k W_k^T-W_k V_k^T
\end{equation}
where $V_k$ and $W_k$ are $n\times k$-matrices. For $k=1$,
$V_1=\bv_1$ and $W_1=\bw_1$. The $k+1$-th update
can then be written as
\begin{equation}
\begin{split}
A^{(k+1)}&=A^{(k)}+\bv_{k+1} \bw_{k+1}^T-\bw_{k+1} \bv_{k+1}^T\\
&=A+V_k W_k^T-W_k V_k^T+\bv_{k+1} \bw_{k+1}^T-\bw_{k+1} \bv_{k+1}^T\\
&=A+V_{k+1} W_{k+1}^T- W_{k+1}^T V_{k+1}^T\,,
\end{split}
\end{equation}
where $V_{k+1}=(V_k,\bv_{k+1})$ and $W_{k+1}=(W_k,\bw_{k+1})$,
and
\begin{equation}
\begin{split}
\bw_{k+1}&=A^{(k)} \bu_{k+1}\\
&=A\bu_{k+1} +V_k W_k^T \bu_{k+1} - W_k V_k^T \bu_{k+1}
\end{split}
\end{equation}
can also be computed without forming $A^{(k)}$ explicitly.

Of course, while it may not be necessary to compute the full $A^{(k)}$
explicitly, the determination of the vectors $\bv_{k+1}$ and
$\bu_{k+1}$ requires the knowledge of the $k+1$-th row or column of
$A^{(k)}$. In practice, the matrix $A$ is therefore partitioned
into panels of $r$ rows and columns. $r$ successive Householder reflections
are then computed and applied one by one to the $r$ rows and columns in the
current panel, but not to the remaining matrix. Instead, they are accumulated
as described above and the remaining part of the matrix is updated in one
block update of the form \eqref{blockhouseeq} which is rich in level-3
matrix operations.

In the case of the Parlett-Reid algorithm, $\bu_k$ is a unit vector
and hence the computation of $\bw_k$ has little cost. In this case
the computational cost is dominated by the outer product update and
hence the block version consists almost entirely of level-3 matrix
operations. This is not the case for the Householder tridiagonalization
where the matrix-vector multiplication to compute $\bw_k$ remains
inherently a level-2 matrix operation.

\section{Upper versus lower triangle storage in the Fortran implementation}
\label{upper_vs_lower}

In order to make full use of the skew-symmetry of the problem, it is
essential that an algorithm only works with either the lower or upper
triangle of the matrix. This is done in the Fortran implementation.
However, in this case it is also mandatory to use mainly column instead
of row operations, as Fortran matrices are stored contiguously in memory
column-by-column. For this reason, the Fortran code implements the
algorithms described in this paper verbatimly only if the lower triangle
of the matrix is provided. Below we briefly describe the differences
when the upper triangle is used.

Instead of starting the tridiagonalization in the first column of the
matrix, the versions using the upper triangle start in the last column.

If a partial tridiagonalization is computed, it is not of the
form \eqref{partialT}, but has $t_{ij}=t_{ji}=0$ only for $i$ even and $j<i-1$.
This amounts to interchanging rows and columns $1$ and $n$, $2$ and $n-1$,
$\dots$, $n/2$ and $n/2+1$ in Eq. \eqref{partialT}. However, since the
determinant of this permutation is equal to one, the value of the
Pfaffian does not change.

In the case of the Parlett-Reid algorithm, a $UTU^T$ decomposition
is computed, where $U$ is an upper unit triangular matrix.


\end{document}